\documentclass[12pt]{article}
\usepackage{graphicx}
\usepackage{float}
\usepackage{amsmath}
\pdfoutput=1

\def\bnl{Physics Department, Brookhaven National Laboratory}

\def\Title#1{\begin{center} {\Large #1 } \end{center}}
\def\Author#1{\begin{center}{ \sc #1} \end{center}}
\def\Address#1{\begin{center}{ \it #1} \end{center}}

\newenvironment{Abstract}{\begin{quotation}  }{\end{quotation}}
\newenvironment{Presented}{\begin{quotation} \begin{center} 
             PRESENTED AT\end{center}\bigskip 
      \begin{center}\begin{large}}{\end{large}\end{center} \end{quotation}}

\def\beq{\begin{equation}}
\def\eeq#1{\label{#1}\end{equation}}
\def\eeqn{\end{equation}}

\def\beqa{\begin{eqnarray}}
\def\eeqa#1{\label{#1}\end{eqnarray}}
\def\eeqan{\end{eqnarray}}

\let\bar=\overbar

\def\Dslash{\not{\hbox{\kern-4pt $D$}}}
\def\dslash{\not{\hbox{\kern-2pt $\del$}}}

\def\msb{{\bar{\ssstyle M \kern -1pt S}}}

\begin{document}
\begin{titlepage}

\vfill
\Title{Longitudinal Spin Physics at RHIC and a Future eRHIC}
\vfill
\Author{ Brian Page}
\Address{\bnl}
\vfill
\begin{Abstract}
\noindent Recent highlights from the spin program at the Relativistic 
Heavy Ion 
Collider (RHIC), focusing on the gluon contribution to the proton spin and 
the polarization of the light flavor sea, are presented. The impact of these 
data on recent global fits by the DSSV and NNPDF groups is also discussed.
Finally, this note examines the constraints on the 
longitudinal spin structure of the proton which would be possible with a
proposed eRHIC facility.

%
\end{Abstract}
\vfill
\begin{Presented}
Conference on the Intersections of Particle\\ 
and Nuclear Physics (CIPANP)\\
Vail, Colorado, USA\\  
May 19--24, 2015
\end{Presented}
\vfill
\end{titlepage}
\def\thefootnote{\fnsymbol{footnote}}
\setcounter{footnote}{0}

\section{Introduction}

Although Quantum Chromodynamics (QCD) has been the accepted theory of strong 
interactions for over forty years, many of its aspects remain poorly 
understood. For example, it is still not clear how the dynamic interactions 
between the fundamental units of QCD, quarks and gluons, give rise to the 
observed spin of the proton. There is a similar lack of understanding about 
the mechanism which is responsible for the generation of the `sea' of light 
flavor quarks and anti-quarks inside the proton. These puzzles have 
motivated a number of theoretical and experimental efforts, including  the
spin program at the Relativistic Heavy Ion Collider (RHIC).

$\quad$

These proceedings will detail recent results from the two main 
experiments 
at RHIC, STAR and PHENIX, which give insight on the puzzles presented 
above as well as present future plans. Section 2 focuses on the latest 
constraints from RHIC on the 
intrinsic gluon spin contribution to the proton's helicity and section 3 
details recent $W$ boson results, which provide a unique probe of the 
polarized sea quark distributions. Section 4 briefly discusses the impact 
that a future upgrade to RHIC (eRHIC), in which one of the proton/ion 
rings is
replaced with an electron ring, could have on our understanding of 
the proton spin structure. Finally, a summary will be given in Section 5.

\section{Gluon Contribution to the Proton Spin}

The proton spin can be decomposed into contributions from the quark and gluon 
helicities and the orbital angular momenta of these partons:

\begin{equation} \label{eq:spinSum}
<S_P> = \frac{1}{2} = \frac{1}{2} \Delta \Sigma + \Delta G + L 
\end{equation}
\begin{equation*}
\Delta \Sigma = \int_0^1 (\Delta u + \Delta d + \Delta \bar{u} + \Delta \bar{d} + \ldots) dx 
\end{equation*}
\begin{equation*}
\Delta G = \int_0^1 \Delta g(x,Q^2) dx
\end{equation*}

\noindent where $\Delta \Sigma$ is the contribution from the spins of the 
quarks and anti-quarks, $\Delta G$ is the contribution from the gluon spin, 
and $L$ is the contribution from angular momentum from the partons.

$\quad$

Fixed-target polarized deep inelastic scattering (DIS) experiments have 
shown that quark
and anti-quark helicities contribute roughly 30\% to the spin of 
the proton in the range $0.001 \leq x \leq 1.0$ \cite{deFlorian:2008mr}. 
This finding naturally lead to the question of how the remaining sources 
of angular momentum, the gluon helicity and parton orbital angular momenta, 
contribute to the proton spin. The orbital angular momenta of the various 
partons is difficult to access experimentally and won't be discussed further, 
but information about the gluon helicity contribution can be obtained 
by measuring scaling violations of the polarized structure function 
$g_{1}(x,Q^2)$ which is accessible in polarized DIS. Because these 
scaling violations are logarithmic in $Q^2$, a large range is needed to 
adequately constrain the gluon helicity contribution. Unfortunately, the 
$Q^2$ range of current polarized DIS data is not large enough to 
significantly constrain the gluon helicity contribution to the proton spin.

$\quad$

The need for better constraints on $\Delta G$ was a primary 
motivation for the spin program at RHIC. Proton-proton collisions probe 
gluon information at leading order via quark-gluon and gluon-gluon hard 
scattering, and because RHIC can provide polarized proton beams, information 
on the gluon polarization is accessible. The observable sensitive to 
$\Delta G$ in polarized $pp$ collisions is the longitudinal double spin 
asymmetry 
$A_{LL}$ where the $LL$ signifies that both beams are longitudinally 
polarized. $A_{LL}$ is defined as: 

\begin{equation} \label{eq:aLLeq}
A_{LL} = \frac{\sigma^{++} - \sigma^{+-}}{\sigma^{++} + \sigma^{+-}}
\end{equation}

\noindent where $\sigma^{++}$ and $\sigma^{+-}$ represent the cross sections 
for some observable when the protons have the same or opposite helicity, 
respectively. The two main experiments at RHIC, STAR and PHENIX, have 
measured 
$A_{LL}$ using a number of different final states, but the most precise 
results are from inclusive jets at STAR and inclusive $\pi^0$'s at PHENIX.

$\quad$

\begin{figure}[h!]
\centering
\includegraphics[keepaspectratio=true, width=0.9\linewidth]{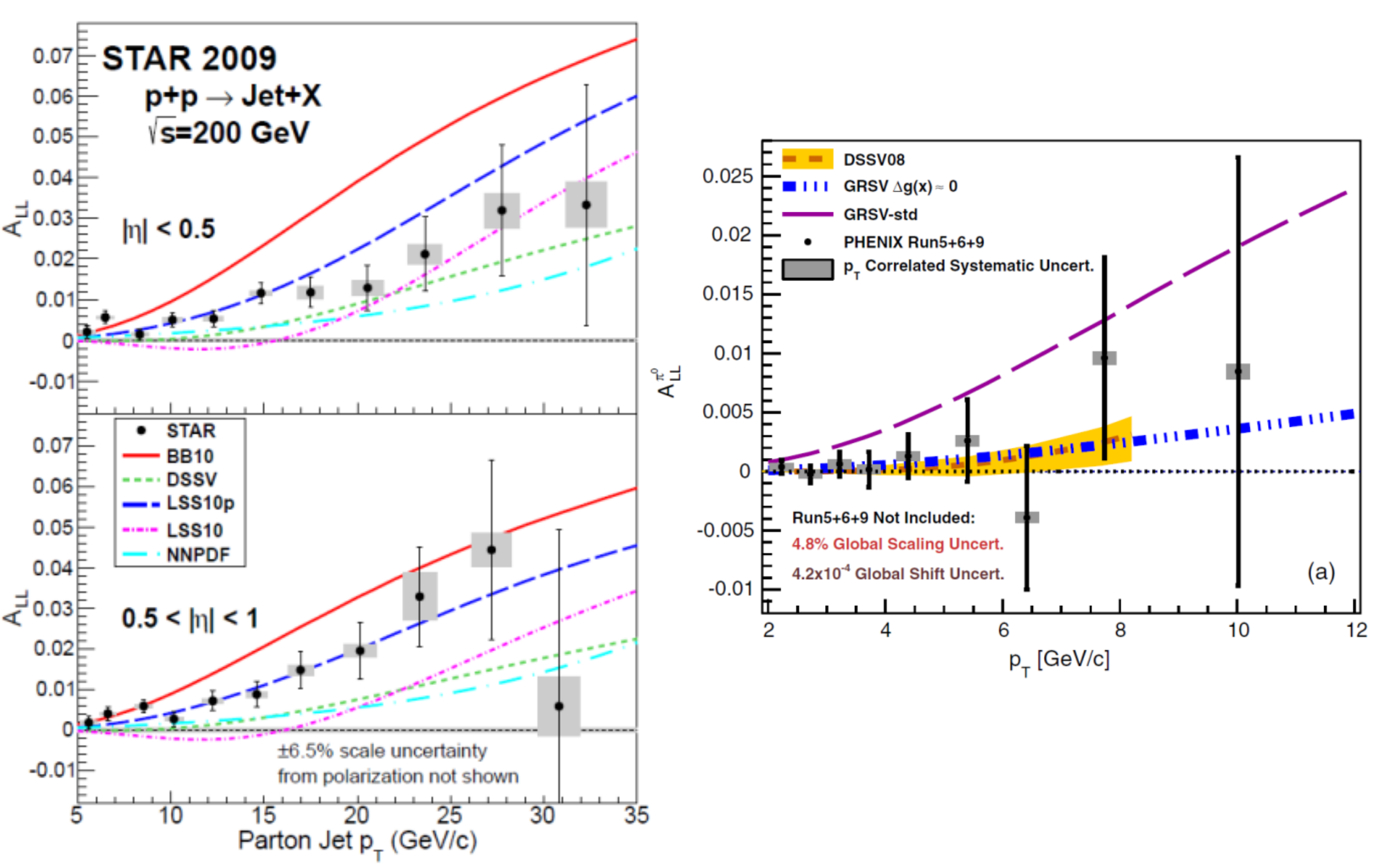}
\caption{(Left) STAR inclusive jet $A_{LL}$ vs parton jet $p_{T}$ for mid-rapidity (upper panel) and forward rapidity (lower panel) jets. (Right) PHENIX inclusive $\pi^0 A_{LL}$ (right) vs pion $p_{T}$ for the combined 2005, 2006, and 2009 data sets. Grey boxes show systematic uncertainties (uncorrelated year-to-year contribution for $\pi^0$) and error bars are statistical. The $\pi^0$ year-to-year correlated systematic uncertainties are shown in the legend. Results are compared to the central values of several global analyses (error bands omitted for clarity).}
\label{fig:starPhenixALL}
\end{figure}

Both collaborations have recently released results from polarized $pp$ data 
taken in 2009 at $\sqrt{s} = 200$ ~GeV 
\cite{Adamczyk:2014ozi,Adare:2014hsq}. The inclusive jet 
$A_{LL}$ from STAR is shown in the left panel of figure 
\ref{fig:starPhenixALL} as a function of 
parton jet $p_{T}$ for two pseudorapidity ranges, which emphasize different 
partonic kinematics. Due to an increase in 
sampled integrated luminosity as well as several advancements in triggering, 
data acquisition, and analysis techniques, these data are a factor of 3 to 4 
more precise than the previous results from the 2006 data set 
\cite{Adamczyk:2012qj}. These results are 
compared to a number of next-to-leading order (NLO) global analyses and lie 
above the original DSSV extraction \cite{deFlorian:2008mr}, which included 
RHIC results from the 2005 and 2006 runs.

$\quad$


The latest $\pi^0 A_{LL}$ results from the PHENIX Collaboration 
are shown in the right panel of figure \ref{fig:starPhenixALL} and combine 
data taken in 2005, 2006, and 2009. The results are again compared to the 
central values of a number of NLO global analyses 
and found to be consistent with the GRSV-zero \cite{Gluck:2000dy} and 
original 
DSSV scenarios. 
These data are also consistent with the STAR results presented above.

$\quad$

\begin{figure}[h!]
\centering
\includegraphics[keepaspectratio=true, width=0.5\linewidth]{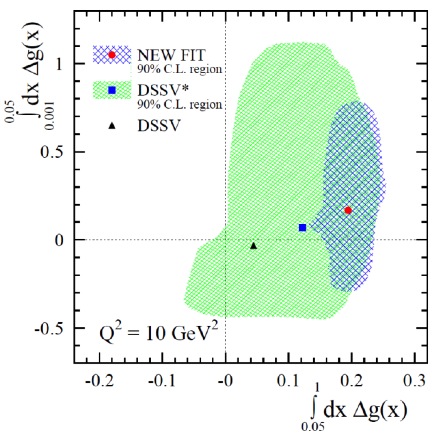}
\caption{Best fit values of the integral of 
$\Delta g(x,Q^{2}=10 \mathrm{GeV}^2)$ for the original and new DSSV extractions (see \cite{deFlorian:2014yva} for details of the data sets). The x-axis is the integral over $0.05 \leq x \leq 1.0$ and the y-axis is over $0.001 \leq x \leq 0.05$.}
\label{fig:dssvResult}
\end{figure}

While the $A_{LL}$ defined in equation \ref{eq:aLLeq} is sensitive 
to $\Delta G$, there is not a direct correspondence due to the convolution of 
different subprocess contributions and probed gluon momentum fractions, both 
of which depend on the transverse momentum of the final state. It is 
therefore necessary to include the RHIC $A_{LL}$ data in an NLO 
`global analysis' which can treat these effects consistently in the 
extraction of $\Delta G$. The DSSV collaboration was the first group 
to carry out a global analysis which incorporated $pp$ data from RHIC as well 
as the world's DIS and semi-inclusive DIS data. Their first results, which 
included RHIC data 
taken in 2005 and 2006, gave a value of $\Delta g(x,Q^2)$ in the range 
$0.05 \leq x \leq 0.2$ equal to $0.005^{+0.129}_{-0.164}$
\cite{deFlorian:2008mr,deFlorian:2009vb}. Recently, the DSSV 
group performed a new analysis incorporating the 2009 jet and $\pi^0$ 
$A_{LL}$ results from STAR and PHENIX and found the first non-zero value for 
$\Delta g(x,Q^2)$. The new result is shown in 
figure \ref{fig:dssvResult} 
and the best fit value for $\Delta g(x,Q^2)$ is $0.20^{+0.06}_{-0.07}$ in the 
range $0.05 \leq x \leq 1.0$ \cite{deFlorian:2014yva}. It is also apparent that 
while the uncertainties  
for $x \geq 0.05$ have improved significantly, the $x$ range below $0.05$ is 
still largely unconstrained. 

$\quad$

The NNPDF collaboration has also begun to incorporate RHIC data into their 
global analysis. The NNPDF technique differs from most other global 
analyses in that it uses a flexible neural network to determine the parton 
distribution functions (PDFs) as opposed to a fixed functional form. Recently, 
a new polarized PDF set was created which includes the STAR jet and $W$ 
data. This set was generated using a re-weighting technique which allows new 
data to be added to an existing NNPDF analysis without having to rerun the 
entire neural network procedure \cite{Nocera:2014gqa}. 
The new polarized PDF set gives a 
value for the integral of $\Delta g(x,Q^2)$ equal to $0.23 \pm 0.06$ in 
the range 
$0.05 \leq x \leq 1.0$, fully consistent with the 
DSSV result.

\section{Sea Quark Polarization}

While the sum over the quark and anti-quark helicity distributions, 
$\Delta \Sigma$, is most 
important for the spin sum rule in equation \ref{eq:spinSum}, the flavor 
separated helicity distributions can also provide important information about 
proton structure and QCD. The 
polarized anti-quark distributions are of particular interest 
as they can discriminate between several different models describing the 
generation of the proton sea and give insight into non-perturbative aspects 
of QCD. Until 
recently, most constraints on the flavor separated polarized quark and 
anti-quark PDFs have come from semi-inclusive DIS (SIDIS), in which an 
identified 
hadron is detected along with the scattered electron. The uncertainty on the 
flavor separated polarized PDFs comes largely from the uncertainty on the 
fragmentation functions, which encode the probability of a quark or 
anti-quark of a given flavor to fragment into a specific hadron, as well as 
the limited $x$ coverage of the SIDIS data. 

$\quad$

The production of real $W$ bosons in polarized $pp$ collisions at RHIC 
provides a 
probe of the flavor separated quark and anti-quark polarized PDFs which is 
complementary to the SIDIS measurements as they do not rely on 
fragmentation functions and probe much higher $Q^2$. At RHIC, $W$ bosons are 
produced primarily via $u + \bar{d}$ and $d + \bar{u}$ s-channel scattering 
and are detected in the charged lepton plus neutrino decay channel 
(where only the charged lepton is measured). The polarized PDFs are accessed 
by measuring a longitudinal single-spin asymmetry $A_L$:

\begin{equation} \label{eq:aLeq}
A_L = \frac{\sigma^{+} - \sigma^{-}}{\sigma^{+} + \sigma^{-}}
\end{equation}

\noindent where $\sigma^{+}$ and $\sigma^{-}$ are the cross sections 
for $W$ production from collisions where the polarized proton beam had 
positive or negative helicity, 
respectively (the spin orientation of the other beam is averaged over). The 
$A_L$ is plotted as a function of the pseudorapidity of the emitted charged 
lepton and different pseudorapidity regions are more sensitive individual 
flavors of quark and anti-quark.


$\quad$

\begin{figure}[h!]
\centering
\includegraphics[keepaspectratio=true, width=0.5\linewidth]{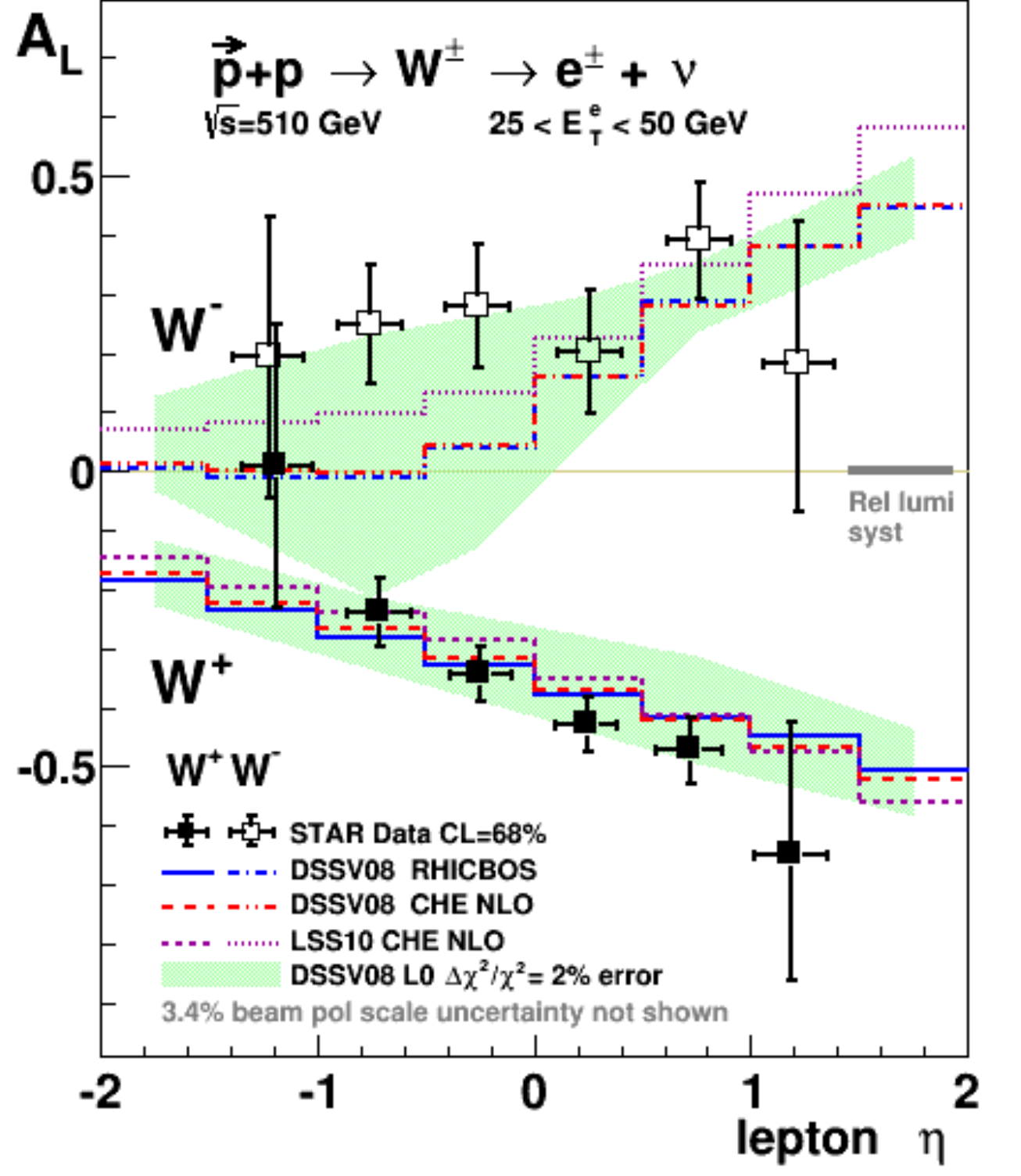}
\caption{$A_L$ from $W^{\pm}$ production plotted as a function of the decay lepton pseudorapidity compared with several theoretical models. Data from 2011 and 2012 were combined using a profile likelihood method and error bars on the points represent the 68\% confidence intervals.}
\label{fig:starWResult}
\end{figure}

The STAR collaboration has recently released $W^{\pm} A_L$ results for data 
taken in 2011 and 2012 \cite{Adamczyk:2014xyw}. The asymmetry, shown in 
figure \ref{fig:starWResult}, 
is presented as a function of lepton pseudorapidity in six bins and compared 
to several theoretical models. The $W^+$ asymmetries agree well with the 
theoretical models but there is some tension between data and theory for the 
$W^-$ asymmetries at negative lepton pseudorapidities. This region has 
enhanced sensitivity to the polarized $\bar{u}$ distributions. 

$\quad$

The PHENIX collaboration has released preliminary $A_L$ results for the 
process $W^{\pm} + Z^0 \rightarrow e^{\pm} + \nu$ in the pseudorapidity range  
$\eta < |0.35|$ from years 2011, 2012, and 2013 \cite{Adare:2015gsd}. These 
data agree well with 
the recent STAR results presented above. In addition to the mid-rapidity 
results, PHENIX has also 
presented asymmetries at larger pseudorapidities than the current STAR results 
using the $W$ to muon decay channel from data taken in 2013. Work is 
continuing on reducing the systematic uncertainties for these results.


$\quad$

\begin{figure}[h!]
\centering
\includegraphics[keepaspectratio=true, width=0.7\linewidth]{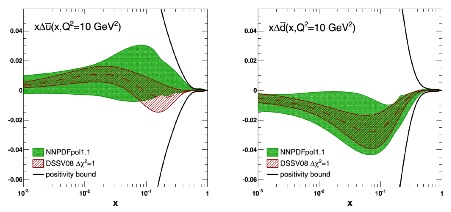}
\caption{Comparison of the most recent polarized NNPDF and original DSSV PDF sets.}
\label{fig:nnpdfWResult}
\end{figure}

As with the extraction of $\Delta G$ from double spin asymmetries, 
constraints on individual quark and anti-quark PDFs from $W$ data 
require a global analysis. Both the DSSV and NNPDF groups have included the 
most recent STAR $W$ results in their latest global analyses 
\cite{Nocera:2014gqa}. A comparison of the original DSSV (which did not 
contain 
RHIC $W$ data) and the latest NNPDF extractions of the polarized light 
anti-quark PDFs can be seen in figure \ref{fig:nnpdfWResult}. As expected, 
the addition of the new STAR $W$ results significantly changes the 
$\Delta \bar{u}$ distribution while only slightly changing $\Delta \bar{d}$.

\section{Possibilities at a Future eRHIC}

As the results in the preceding sections show, the RHIC Spin Program has 
yielded a wealth of information about the polarized structure of the proton. 
While $pp$ scattering has the advantage of accessing gluon information at 
leading order, the fact that both beam and target are composite objects means 
that precise determination of the partonic kinematics is very difficult. This 
increases the uncertainty on the $x$ dependence of the extracted polarized 
PDFs. Because a lepton has no internal structure, the partonic kinematics can 
be accessed much more precisely in DIS. As mentioned above, fixed target DIS 
experiments do not have the kinematic reach needed to constrain $\Delta G$, 
but a much wider kinematic range would be available at a polarized $ep$ 
collider. There is a great deal of interest among the nuclear physics 
community for building such a machine and one proposed realization would 
consist of replacing one of the current RHIC rings with an electron ring to 
form an eRHIC \cite{Aschenauer:2014cki}.

$\quad$

An eRHIC would allow for collisions of electrons and polarized protons, as 
well as heavier nuclei. Such a machine would marry the wide kinematic range 
possible in colliders with the precision of DIS and would be the ultimate 
laboratory for the exploration of QCD. A future eRHIC will have a large and 
diverse scientific mission \cite{Accardi:2012qut}, one part of which will 
be the precise determination of $\Delta G$ and the quark and anti-quark 
PDFs discussed in the sections above. 

$\quad$

\begin{figure}[h!]
\centering
\includegraphics[keepaspectratio=true, width=0.9\linewidth]{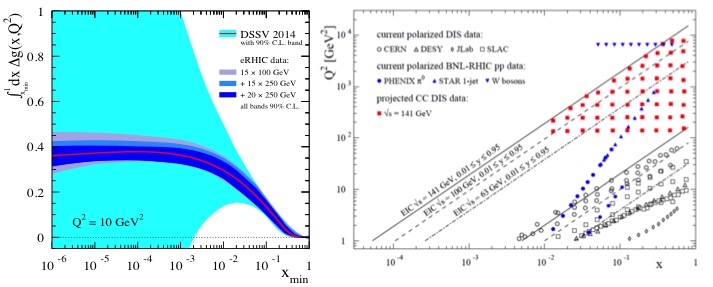}
\caption{(Left) Running integral of $\Delta g(x,Q^2)$ from $x_{\mathrm{Min}}$ to unity as a function of $x_{\mathrm{Min}}$. The light blue region represents the uncertainties from the most recent DSSV extraction while the darker blue regions show the expected uncertainties with eRHIC data. (Right) Expected $x-Q^2$ coverage of an eRHIC compared with coverage of existing polarized data sets. The red squares show the expected coverage of eRHIC charged-current data.}
\label{fig:eRHIC}
\end{figure}

The golden channel for measuring $\Delta G$ at an eRHIC will be scaling 
violations of the polarized structure function $g_1(x,Q^2)$. The collider will 
allow a much wider lever arm in $Q^2$ for a given value of $x$ than the 
current fixed target DIS experiments. In addition, the proposed high luminosity 
will greatly improve the statistical error on each point. The left panel of 
figure \ref{fig:eRHIC} presents the estimated precision on the integral of 
$\Delta g(x,Q^2)$ between $x_{\mathrm{Min}}$ and unity as a function of 
$x_{\mathrm{Min}}$ for several expected eRHIC energies and integrated 
luminosities. The uncertainties from the most recent DSSV extraction are shown 
for comparison and the power of eRHIC to improve the constraints on $\Delta G$ 
is obvious.

$\quad$

The proposed eRHIC machine will also be able to probe the flavor separated 
(un)polarized quark and anti-quark PDFs in two complimentary ways: SIDIS and 
virtual $W$ production 
\cite{Aschenauer:2013iia,Aschenauer:2014cki,Accardi:2012qut}. As any eRHIC 
detector is likely to have particle 
identification abilities, the standard SIDIS measurements currently carried 
out at fixed target experiments will be available at an eRHIC. These 
measurements will benefit from improved fragmentation function determinations 
which should be available by the time an eRHIC turns on. An eRHIC should  
contribute to the improvement of these fragmentation functions as well. 
The second method for 
accessing the quark and anti-quark PDFs involves charged-current DIS in 
which a virtual $W$ boson is exchanged instead of the usual photon. These 
charged-current events access different polarized structure functions 
which encode unique combinations of quark and anti-quark PDFs. In addition, 
because the $W$s are virtual, a range of $Q^2$ values can be probed and an 
overlap between the $W$ and SIDIS measurements will be possible. The right 
panel of figure \ref{fig:eRHIC} shows the expected $x-Q^2$ coverage of the 
charged-current measurements in the eRHIC kinematic range. As with the 
extraction of $\Delta G$, the large kinematic range and high statistics 
possible with an eRHIC is expected to greatly reduce the uncertainty on the 
flavor separated quark and anti-quark PDFs.

\section{Summary}

The RHIC Spin Program has given great insight into the polarized structure of 
the proton. Recent jet and pion asymmetries have been incorporated into NLO 
global analyses and 
give evidence for the first non-zero gluon contribution to the spin of the 
proton, albeit over a restricted $x$ range. The $W$ program has also yielded 
tighter constraints on the polarized quark and anti-quark PDFs, having a 
particularly significant effect on the $\Delta \bar{u}$ distribution. There are 
plans to build a polarized electron-ion collider which would become the 
premier facility for the exploration of QCD. Such a machine would allow 
precision determinations of both the gluon contribution to the proton spin and 
the flavor separated quark and anti-quark distribution functions.


\end{document}